\documentclass[apl,superscriptaddress,reprint,amssymb,aps,floatfix,showkeys]{revtex4-1}
\usepackage{amsmath}
\usepackage{amsfonts}
\usepackage{amssymb}
\usepackage{graphicx}
\usepackage{color}
\usepackage{tabularx}
\usepackage{braket}
\usepackage{url}
\usepackage{hyperref}

\begin{document}
\title{A Search for Effects of Cosmic Rays with Multi-scale Entropy Metrics}

\author{William M. Campbell}
\affiliation{National Quantum Computing Testbed, School of Mathematics and Physics, University of Queensland, St Lucia, QLD, 4072, Australia}
\affiliation{Quantum Technologies and Dark Matter Labs, Department of Physics, University of Western Australia, Crawley, WA, 6009, Australia}

\author{Ben T. McAllister}
\affiliation{Centre for Astrophysics and Supercomputing, Swinburne University of Technology, Hawthorn VIC 3122, Australia}
\affiliation{Quantum Technologies and Dark Matter Labs, Department of Physics, University of Western Australia, Crawley, WA, 6009, Australia}

\author{Eugene N. Ivanov}
\affiliation{Quantum Technologies and Dark Matter Labs, Department of Physics, University of Western Australia, Crawley, WA, 6009, Australia}

\author{Michael E. Tobar}
\affiliation{Quantum Technologies and Dark Matter Labs, Department of Physics, University of Western Australia, Crawley, WA, 6009, Australia}

\author{Mehran Mossammaparast}
\affiliation{Wenzel Associates Inc. Austin, TX, USA}

\author{Mike Sawicki}
\affiliation{Wenzel Associates Inc. Austin, TX, USA}

\author{Maxim Goryachev}
\email{maxim.goryachev@uwa.edu.au}
\affiliation{Quantum Technologies and Dark Matter Labs, Department of Physics, University of Western Australia, Crawley, WA, 6009, Australia}

\begin{abstract}

%We report the comparison of frequency fluctuations from bulk acoustic wave oscillators, measured both above and one kilometre below the ground, in a low muon background radiation environment. These tests are motivated by the potential contribution of ionizing-radiation backgrounds to transient perturbations in mechanical resonant sensors. Standard frequency-stability and spectral analysis techniques show no statistically compelling separation between the two environments over the explored time scales. In contrast, multi-scale sample entropy metrics reveal a pronounced divergence at most time scales. Such deviation is indicative of an environment-dependent change in the predictability of oscillator fluctuations. By indicating  evidence for radiation impact on mechanical oscillators, These results demonstrate that entropy-based measurement metrics provide information complementary to Allan deviation and power spectral density methods, with enhanced sensitivity to intermittent structure and non-Gaussian contributions that can be diluted in conventional metrological techniques. We propose multi-scale sample entropy as a practical addition to the frequency-metrology toolbox and as a diagnostic relevant to future fundamental tests using cryogenic resonant sensors, where rare-event backgrounds and poorly understood $1/f$ noise can limit performance.

We report a comparison of frequency fluctuations in oven-controlled quartz bulk-acoustic-wave oscillators operated above ground and one kilometre underground in a low-muon-background environment. The experiment is motivated by the possibility that cosmic rays and other ionizing-radiation backgrounds produce rare, impulsive energy-deposition events that perturb high-Q mechanical resonators and appear as intermittent, non-Gaussian structure in oscillator frequency noise. Conventional power spectral density and Allan-deviation analyses show no statistically compelling separation between the two environments over the explored timescales. In contrast, multi-scale sample entropy and its modified form reveal a pronounced divergence, with the underground data exhibiting increased predictability over a broad range of effective integration times. This result identifies a change in the temporal structure of the oscillator fluctuations that is largely hidden from standard second-order frequency-stability metrics. We therefore propose multi-scale sample entropy as a new diagnostic for frequency control and timing, complementary to Allan deviation and spectral analysis, with particular sensitivity to intermittent structure, non-stationary contributions, and rare-event contamination. The observed entropy separation also provides evidence that the above-ground cosmic-ray environment influences oscillator frequency fluctuations, suggesting that radiation-linked disturbances may contribute to the stochastic behaviour of precision mechanical oscillators. These findings introduce an entropy-based methodology for oscillator metrology and provide a practical tool for future fundamental-physics experiments using cryogenic resonant sensors, where rare-event backgrounds and poorly understood low-frequency noise can limit sensitivity.

\end{abstract}
\date{\today}
\maketitle

\section*{Introduction}

Cosmic rays and other ionizing particles are a persistent, disturbance channel in precision measurement systems. The injection of rare, impulsive energy-deposition events can excite long-lived mechanical or electromagnetic responses, produce intermittent transients, and bias measurement metrics with non-stationary statistics. In resonant-mass gravitational wave detectors, such coupling is well studied; the NAUTILUS gravitational-wave bar detector observed measurable responses correlated with extensive air showers, demonstrating that energetic particles can drive macroscopic vibrational excitations in cryogenic resonant bars\cite{Magalhaes:2001aa, Astone:2002ab}. Quartz bulk acoustic wave (BAW) resonators are conceptually analogous, high-Q elastic systems with exceptionally long ringdown times\cite{Goryachev1, quartzPRL, ScRep}, so it is natural to expect that particle interactions may excite vibrational modes and cause perturbative signatures in frequency and phase readout. The same issue has recently emerged as a limiting error channel in superconducting quantum processors, where cosmic-ray muons and environmental radioactivity can generate nonequilibrium quasiparticles and phonon-mediated bursts that produce spatially correlated relaxation and charge-parity errors across many qubits\cite{Martinis:2021aa,osti_1843566,Harrington:2025aa}. This connection highlights that rare particle-induced events are not merely a background in cryogenic precision systems, but a common impulsive-noise mechanism relevant to both high-Q mechanical resonators and fault-tolerant superconducting quantum computing.

Such high sensitivity resonant mass sensors are gaining relevance in the context of emerging high-frequency gravitational-wave instrumentation\cite{Goryachev:2014ac,Campbell:2023aa,Campbell:2021aa}, other tests of fundamental physics\cite{Lo:2016aa,Campbell:2023ab,Campbell:2021aa} as well as their conventional applications in precision timing standards and metrology\cite{Salzenstein:2016aa,Liu:2023aa,AA:2024aa}. Multiple experiments for example have recently considered utilising piezoelectric vibrational medium at cryogenic temperatures as transducers for high-frequency strain \cite{Campbell2025, Campbell:2023aa, Chu2026, Albani2025}, targeting beyond standard model sources of gravitational waves\cite{Aggarwal:2021aa}. In these experiments weak excitations must be discriminated from a complex background of instrumental noise. Such detectors are subject to particle-induced disturbances as a plausible background \cite{Goryachev:2021aa}, potentially biasing detection statistics. Rare event perturbations have also been reported across a range of precision metrology platforms, including superconducting qubits \cite{McEwen2022, Bratrud2025} where correlated burst-like signatures have been attributed to ionizing radiation and secondary particle showers. Collectively, these observations motivate further investigation as to whether and how cosmic rays contribute to the stochastic performance and transient behaviour of acoustic oscillators.

In quartz BAW systems specifically, excess low-frequency noise is historically not well understood. Many quartz-based oscillators and resonators exhibit pronounced flicker or $1/f$ noise whose microscopic origin remains debated \cite{Walls:1992aa,Galliou2013a,Sthal:2015aa}, with proposed mechanisms spanning defect dynamics, surface and electrode effects, strain-mediated two-level systems, and coupling to readout and environment. If a fraction of the observed low-frequency noise arises from impulsive processes, whether due to ionizing radiation, microphonic disturbances, or sporadic relaxation events, then conventional stationary-noise models can be misleading. Underground laboratories\cite{Bettini:2011aa} provide a controlled inherent suppression to the dominant cosmic-ray muon flux while keeping much of the apparatus unchanged. Comparing otherwise identical measurements performed above ground and underground therefore offers a direct test of whether rare, radiation-linked disturbances measurably contribute to the observed frequency fluctuations and their statistical structure. To the best of our knowledge, such a controlled comparison both above and below ground has not previously been reported for oscillators operating in a frequency-metrology readout configuration.

In this work, we perform long-duration measurements of quartz BAW oscillator frequency fluctuations in two environments; above ground, and in a one-kilometre-deep underground facility\cite{SUPL1}. We compare the resulting signatures across a broad range of averaging times. In addition to standard frequency-stability analysis, we employ sample entropy and its multi-scale generalisations as metrics to explore the regularity of time series structure. Entropy-based complexity measures are widely used in biomedical and nonlinear-systems analysis to quantify irregularity and loss of predictability \cite{Delgado-Bonal:2019aa,Richman:2000aa,Costa:2005aa}, but they have seen no adoption in frequency metrology, where Allan deviation (ADEV) and related two-sample variances are the prevailing standards. Allan deviation\cite{Allan:1966aa} is however fundamentally a second-order statistic tuned to stationary noise processes and variance scaling with averaging time. Whereas cosmic-ray-like disturbances are intermittent and non-Gaussian, often manifesting as localized transients. Such events can alter pattern statistics of the time series without producing change in the global two-sample variance. Sample entropy, by contrast, measures the rate at which short patterns cease to repeat when extended by one or more samples. It is therefore naturally sensitive to transient structure and changes in predictability, making it a promising diagnostic for rare-event contamination in otherwise stationary frequency series. By combining a controlled environmental comparison in both high and low background environments, with an entropy-based analysis framework, we explore the extent to which rare-event backgrounds contribute to the observed low-frequency behaviour of quartz BAW oscillators, and provide a methodological example for the incorporation of multi-scale entropy into frequency metrology and rare-event studies.

\section{Multi Scale Sample Entropy}

Allan Deviation is a statistical measure used to analyse the stability of frequency signals over different time scales \cite{Allan:1966aa, Rubiola:2008aa}. Unlike the ordinary standard deviation, whose interpretation as a single stability metric is most meaningful for stationary data, the Allan deviation characterises fractional-frequency fluctuations as a function of averaging time. This makes it well suited for oscillators, atomic clocks, and precision timing devices, where noise, drift, and instability often depend on the measurement timescale. It works by computing the root mean square of fractional frequency differences over adjacent averaging intervals, revealing noise characteristics such as white phase and frequency noise, flicker noise, and drift. In frequency control, ADEV helps identify the optimal averaging time to minimise noise effects, improve synchronisation, and enhance the precision of timing systems used in GPS, telecommunications, and scientific instrumentation.

Entropy quantifies uncertainty or randomness in a system and is widely used in complexity analysis, time-series analysis, and information theory. Independent of thermodynamic definitions of entropy, various metrics have been introduced to capture different aspects of statistical disorder. Sample entropy assesses the predictability of a time series by measuring the likelihood that two sequences remain similar in the proceeding step, with lower values indicating higher predictability.

Sample entropy $SampEn(m,r,x)$ is computed as the negative natural logarithm of the conditional probability of $m$ consecutive samples having a maximum separation greater than some predetermined threshold $r$ for some time series $x$. {The exact formulation is described in Appendix~\ref{app:multiscale_sample_entropy} and also in literature \cite{Costa2002, Busa2106, Wu2013, Humeau205}.} This metric can extended  to multiple time scales by determining a coarse-grained version of the original signal, averaging every $M$ consecutive points where the effective time scale is defined as $\tau = M / f_s$, with $f_s$ being the sampling frequency. Following the methods presented in other works, the coarse graining procedure divides a time series $\{x(i):1\leq i\leq N\}$ as
\begin{equation}
	X_\tau = \frac{1}{M}\sum_{j=1}^{M} x(M i - M +j): 1\leq i \leq N,
\end{equation}
sample entropy is then determined for each scale averaged signal $\{SampEn(m,r,X_\tau):1\leq \tau \leq \frac{N}{M}\}$, {where the embedding dimension ($m$) specifies the length of the short temporal patterns compared in the time series. Dimension $m=2$ tests whether pairs of neighbouring samples that are initially similar remain similar when extended to three samples. }For this case, the matching threshold $r$ is usually set to a fraction of the standard deviation of the original time series. Alternatively, one can also recompute this threshold at each scale utilising the standard deviation of each $X_\tau$. This method may make the interpretation of multi scale entropy challenging as changes to the standard deviation will also contain temporal information of the original signal's statistics, resulting in a dynamic definition for entropy. However, as no universal relationship between entropy and standard deviation exists \cite{CostaReply2004} either method can be utilised depending on the exact application. 
 
 We investigate multi scale sample entropy as a tool for analysing frequency fluctuations, making a direct comparison with Allan deviation. In order to investigate the suitability for identifying noise processes in metrological time series, we define the multi scale sample entropy $(S_E)$, which is determined utilising the standard deviation of the original time series $x(t)$, as well as the modified multi scale sample entropy $(\tilde{S}_E)$ which recomputes $r$ at each scale using the standard deviation of $X\tau$. 
 
\section{Sample Entropy for Frequency Fluctuation Analysis}
Noise processes that enter a carrier signal may be characterised by their spectral scaling. Types of noise commonly encountered in frequency control applications include white or stationary noise, $1/f$  or flicker noise,  and $1/f^2$, $1/f^3$ coloured noise, and drift. These noise processes represent different underlying mechanisms affecting the stability of frequency sources, such as oscillators and atomic clocks. White noise corresponds to uncorrelated random fluctuations, often arising from thermal or electronic sources. $1/f$ noise is characterised by long-term correlations and is prevalent in electronic components and resonators. $1/f^2$ noise, often referred to as random walk frequency noise, describes processes with even stronger memory effects, leading to cumulative drift over time. $1/f^3$ noise represents a deeper level of correlation, often associated with long-term instability and environmental perturbations. Drift refers to slow, systematic changes in frequency, which can be caused by ageing, temperature variations, or other external influences. In practical frequency control systems, such as oven-controlled crystal oscillators (OCXOs) \cite{Frerking:1996aa}, all of these noise types can be present simultaneously, with different contributions depending on the time scale of observation. Understanding how these noise processes interact and dominate at different scales is essential for optimising precision timing systems, and serves as a motivation for further analysis using entropy-based methods.

It's worth noting that entropy based metrics cannot replace ADEV as a measure of frequency stability. ADEV specifically characterises variations in frequency over different timescales, making it sensitive to absolute values of input time series. Multiplying a frequency time series by a constant factor directly scales the Allan deviation, either increasing or decreasing it proportionally. In contrast; sample entropy remains invariant under constant multipliers since it measures the complexity and unpredictability of a signal rather than its absolute magnitude. This fundamental difference means that sample entropy captures structural randomness but does not reflect the magnitude of frequency fluctuations, which is essential for evaluating long-term stability in timekeeping, metrology, and signal processing applications. Alternatively however, one should think of entropy measures as a complimentary probe with which the frequency deviation time series can be further studied, potentially revealing trends on various scales that are unable to be identified by ADEV analysis.
 
\begin{figure}
     \begin{center}
            \includegraphics[width=0.47\textwidth]{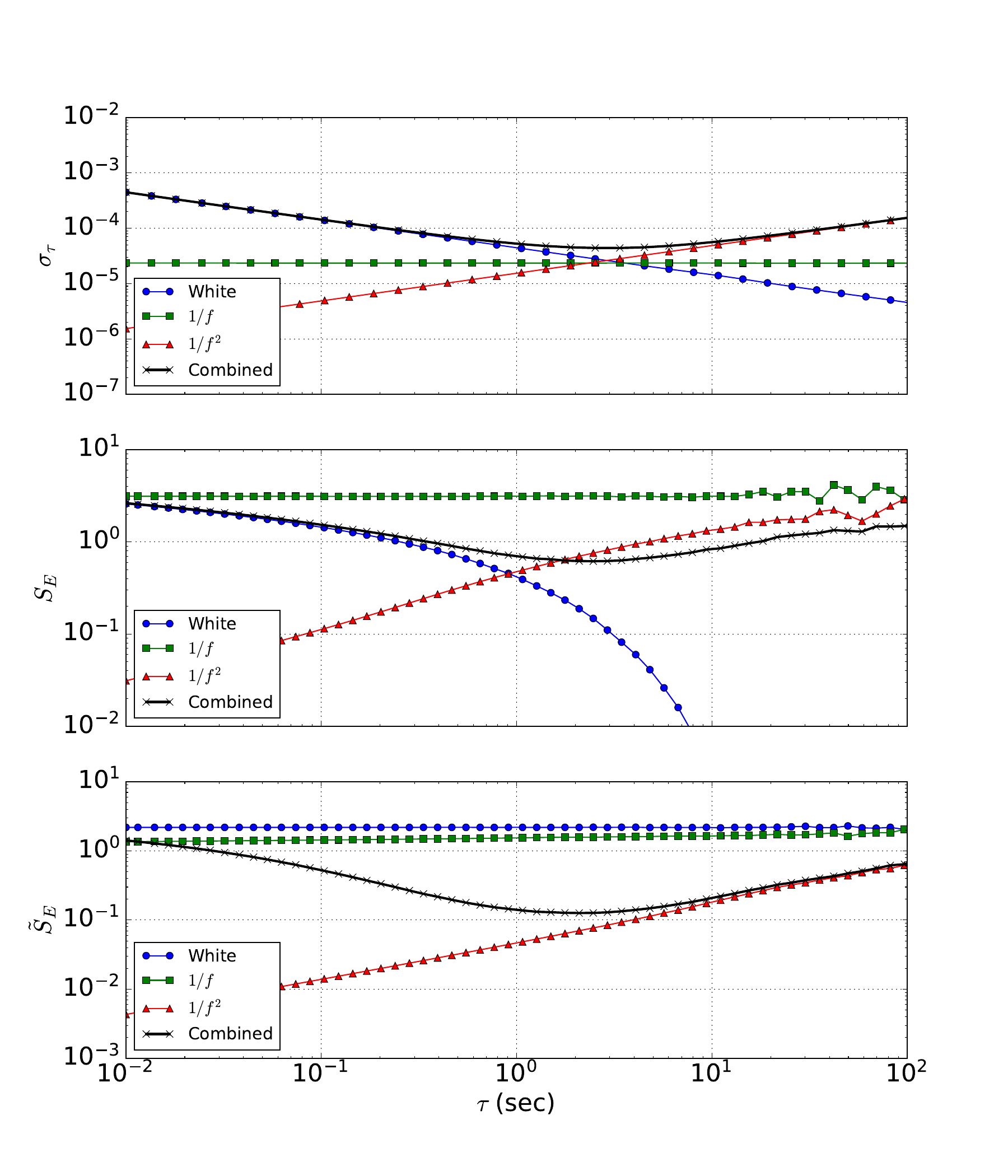}
            \end{center}
    \caption{Comparison Allan deviation a), multi scale sample entropy b) and modified multi scale sample entropy c) for white, flicker and $1/f^2$ noise types as well as the combined noise of a typical oscillator.}%
   \label{OCXO}
\end{figure}

To investigate the characterises of these metrics, an idealistic oscillator noise spectrum is modelled by combining different processes; white noise in the high-frequency range, $1/f$ in the mid-range, and $1/f^2$ noise in the low-frequency region. The total ADEV is then computed for the combined signal, as well as the contributions from individual noise components, to assess frequency stability across all timescales. Multi-scale sample entropy $S_E$ and $\tilde{S}_E$ are also computed for each noise component. Both Allan deviation and sample entropy metrics are plotted in figure \ref{OCXO}, providing a direct comparison between spectral properties and time-domain unpredictability.

\begin{figure}
	\begin{center}
		\includegraphics[width=0.47\textwidth]{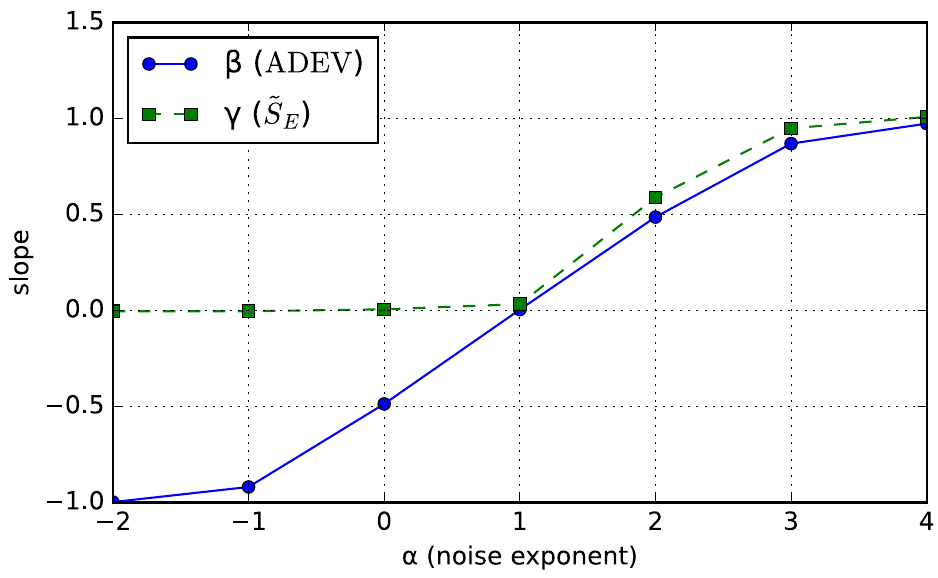}
	\end{center}
	\caption{The logarithmic scaling of $\tilde{S}_E$ and ADEV  determined for the noise process $1/f^\alpha$ are plotted for various $\alpha$. Green markers show $\gamma$ as the scaling exponent of $\tilde{S}_E$ , whilst similarly the blue markers show the scaling exponent of ADEV in $\beta$. }
	\label{fig:exp}
\end{figure}
Fig.~\ref{OCXO} demonstrates how sample entropy effectively distinguishes between different types of noise. In the case of white noise, $S_{E}$ decays for increasing time scales as the time series lacks any long term structure. Coarse-graining at large scales averages out local fluctuations and the Gaussian nature of the time series thus leads to low pattern unpredictability. For flicker noise, the entropy curve is flat across all scales. This is reflective of the long-range correlations and scale invariant structure inherent to flicker noise as its unpredictability is retained equally at all scales. Coloured noise exhibits low entropy at short time scales as the high degree of structure in the time series allows for large predictability between successive samples.  Furthermore, for a composite signal containing multiple noise types, the sample entropy varies across time scales, dynamically reflecting the competing noise processes at each scale. {It should be noted that the apparent fluctuations of $S_E$ at the largest values of $\tau$ in Fig.~\ref{OCXO} are finite-sample artefacts rather than physical instabilities of the simulated noise.}

For $\tilde{S}_E$, the interpretation is less clear as the entropic definition of signal regularity changes for each scale due to the changing standard deviation. However the scale dynamics still contain experimentally useful properties. While the white noise curve is flat, the $1/f^2$ component demonstrates a consistent power scaling $\propto \tau^{0.6}$. In fact $\tilde{S}_E$ can differentiate between noise processes much like ADEV by examining the power law $\tilde{S}_E \propto \tau^\gamma$ for some scaling index $\gamma$. In figure \ref{fig:exp} we plot $\gamma$ extracted from noise process of different spectral shape $1/f^\alpha$. This can then be compared to the same scaling for ADEV $\propto\tau^\beta$ for some index $\beta$. We find for positive values of $\alpha$ both $\tilde{S}_E$ and ADEV can separate out contributing noise processes based on their scaling relationships.
 
\section{Event Based Noise}
\begin{figure}
     \begin{center}
            \includegraphics[width=0.49\textwidth]{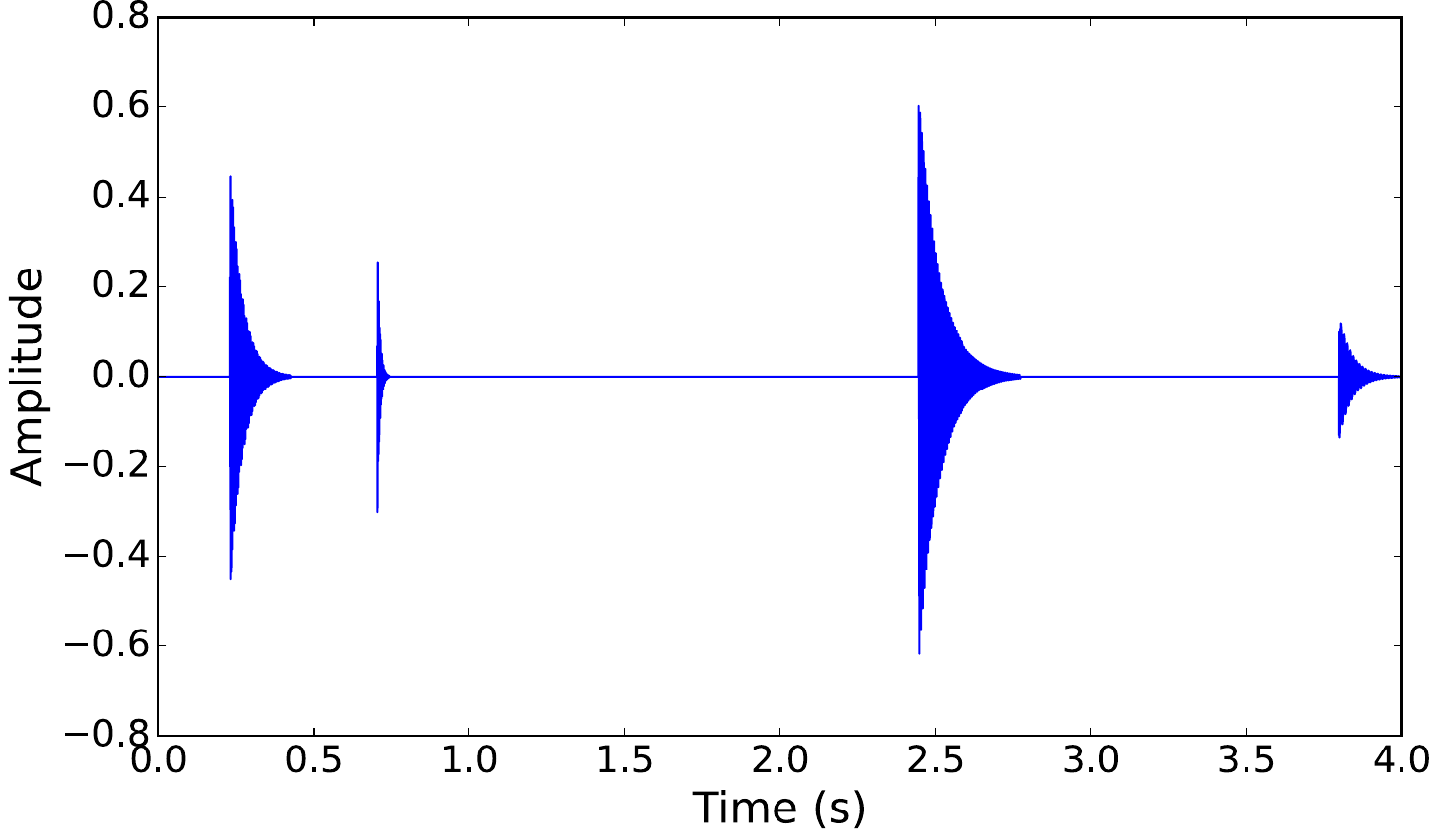}
            \end{center}
    \caption{Time trace signal demonstrating 4 random impulse events in a dilute regime.}%
   \label{events}
\end{figure}

\begin{figure}
     \begin{center}
            \includegraphics[width=0.49\textwidth]{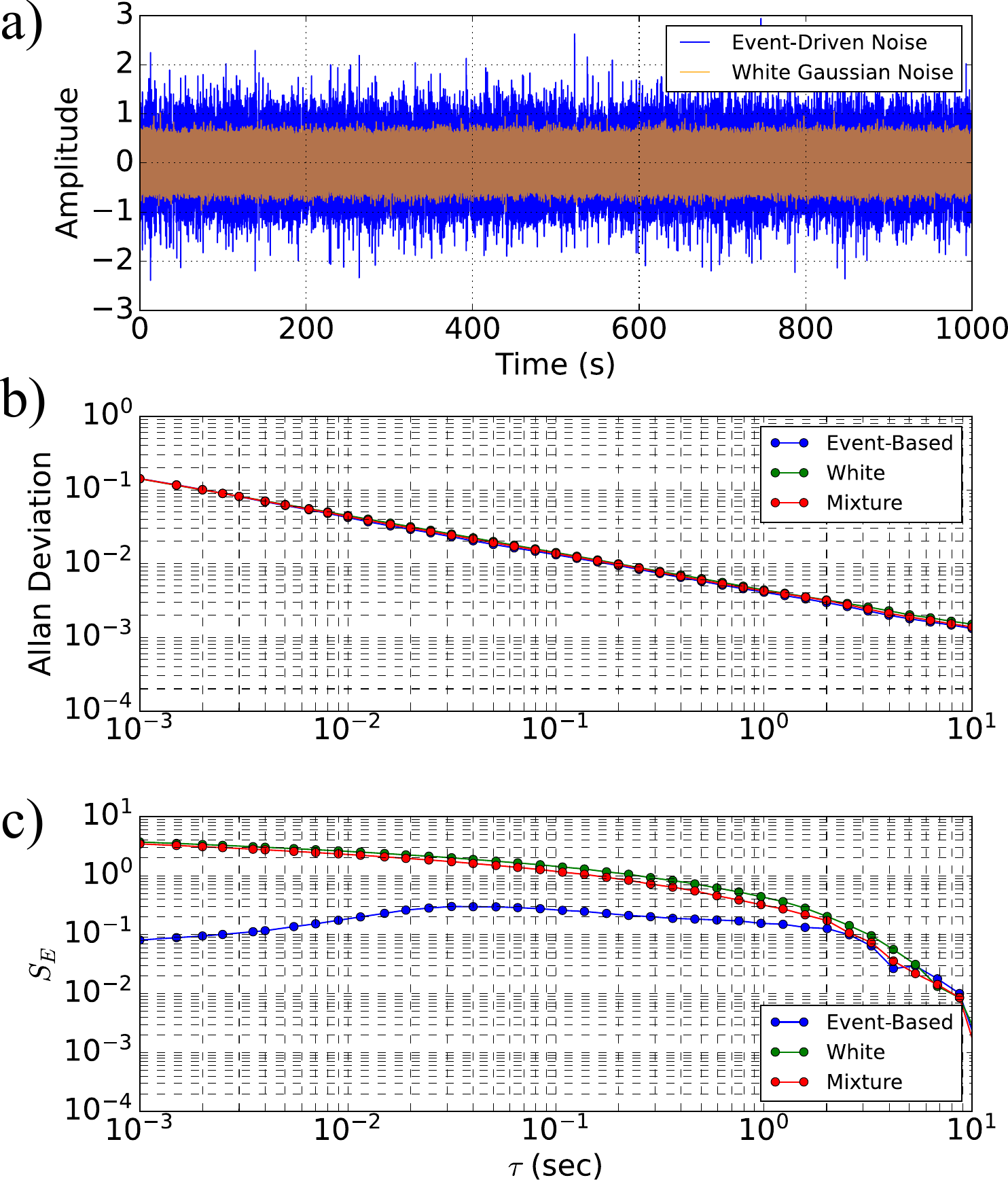}
            \end{center}
    \caption{Analysis of different noise types: a) Time traces of white and event based noise, b) Allan deviation, c) multi scale sample entropy.}%
   \label{eventbase}
\end{figure}

 To demonstrate an application of entropy in analysing frequency fluctuations, we consider the case of noise dominated by transient impulse events. Standard gaussian noise is thus compared against a noise process generated by including random impulse events with a second-order oscillatory decay, introducing a structured, yet stochastic, component to the signal. In this model, multiple parameters are randomised, including the event start times $t_0$, oscillation frequency $f_0$, decay rate $T$, initial phase $\phi$ and amplitude $V$, creating a complex time-dependent fluctuation pattern:
 \begin{equation}
	\label{FF0001G}
v_{\textrm{event}}(t) = V\sin(2 \pi f_0 (t-t_0) + \phi) \exp(-(t-t_0)/T)u(t-t_0)
\end{equation}
where $u (t)$ Heaviside step function. A realisation of such signal with dilute events is shown in Fig.~\ref{events}.
This type of noise can arise in real-world systems due to transient disturbances, relaxation phenomena, or resonant interactions in random sources.
 Figure \ref{eventbase} compares a white noise, event-based noise and a mixed equal parts white and event-based noise process. The time-domain representation, Allan deviation, and multi-scale sample entropy is shown for each type of noise. In the time domain, both noise types appear visually similar, making it difficult to distinguish between them. Likewise, Allan deviation, which primarily captures variance at different time scales, fails to reveal any significant difference between the two signals, as both exhibit similar scaling behaviour. However, when analysed using $S_E$, a clear distinction emerges: while white noise maintains a relatively large entropy profile, the event-based noise exhibits a reduced entropy at short time scales that reflects its underlying structured fluctuations. Modified multi scale entropy $\tilde{S}_E$ also follows the same trend.
In practical scenarios, it is common for both white noise and non-stationary event-based noise to be present simultaneously, leading to a mixed signal with contributions from both sources. To analyse how the distinction between these two noise types evolves as their relative proportions change, $S_E$ and ADEV can be computed for a combined signal defined as
\begin{equation}
	\label{eq:mixed}
v_{\mathrm{mixed}}(t) = \chi v_{\textrm{white}}(t) + (1 - \chi) v_{\textrm{event}}(t),
\end{equation}
where $v$ represents the corresponding signal values as a function of time, and $\chi$ is a mixing parameter that controls the relative weight of white, and event-based noise. {The normalized difference of $v_{\mathrm{mixed}}(t)$ to the case of pure white noise is then
\begin{equation}
	\label{FF0002G}
	\Delta M = \frac{\left\langle |M^{\mathrm{mixed}}(\chi) - M^{\mathrm{white}}| \right\rangle}{\left\langle M^{\mathrm{white}} \right\rangle},
\end{equation}

where $M$ represents either the sample entropy $S_E$ or Allan deviation $\sigma_\tau$, and angle brackets denote averaging across all scales (time scales for $S_E$, integration times for $\sigma_\tau$). Figure \ref{fig:mixing} shows that the difference in sample entropy becomes greater than 3dB of its value in the case of pure white noise for  mixture concentrations where $\chi<0.06$, identifying the point at which event-driven noise becomes detectable. This suggest that sample entropy may be a better metric than ADEV at identifying temporal structured processes buried in white noise.}

\section{Spurious Signals}
Measurements are often corrupted by the presence of spurious signals, which can arise from various sources such as mains harmonics, digitisation artefacts, or signal pickup from unintended electromagnetic interference. These unwanted contributions can distort the true signal, affecting accuracy and making it challenging to extract meaningful information, particularly in precision measurements and frequency stability analysis. 

\begin{figure}
     \begin{center}
            \includegraphics[width=0.46\textwidth]{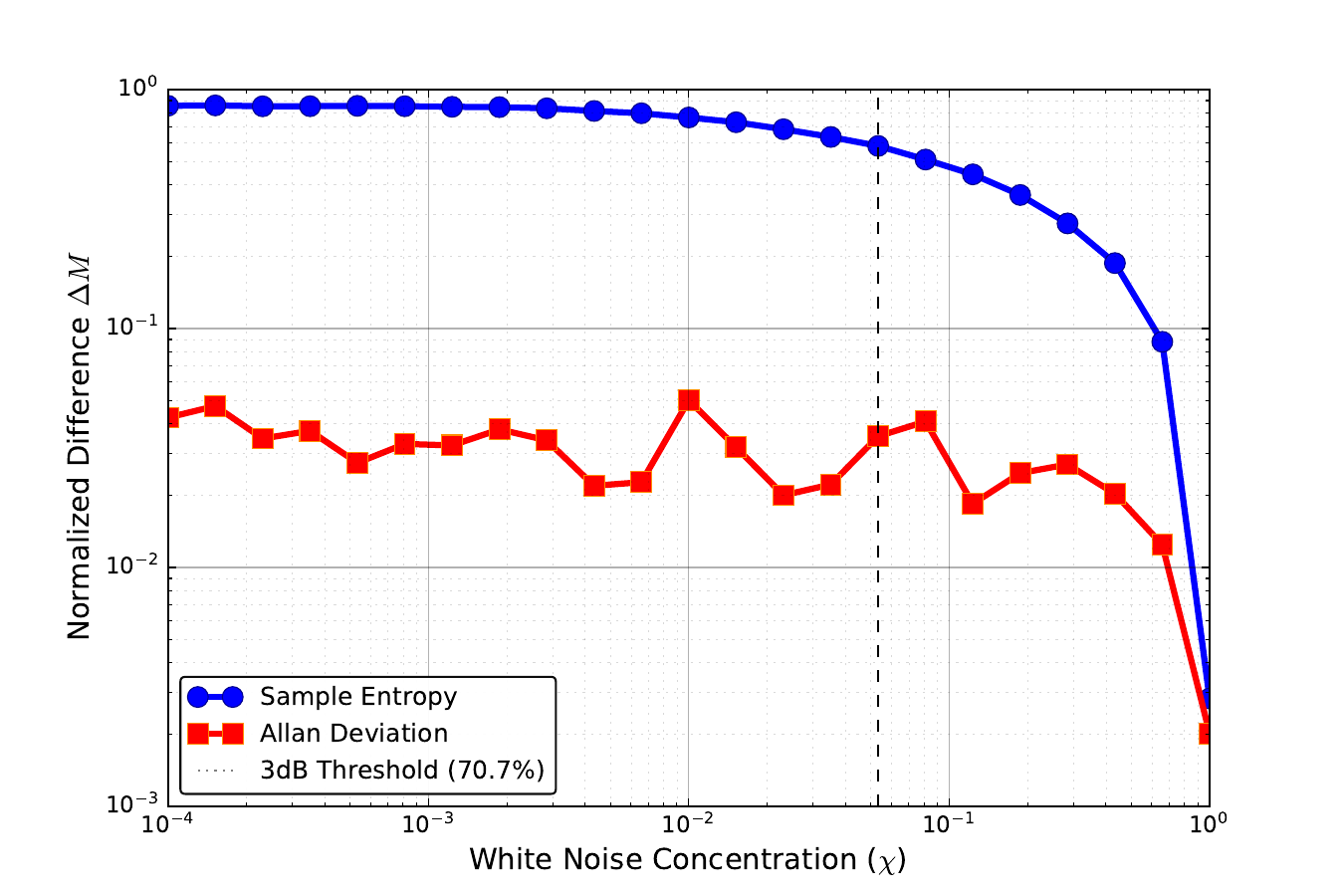}
            \end{center}
    \caption{\label{fig:mixing}Deviation of sample entropy from that of white noise and a harmonic signal as a function of mixing parameter $\chi$.}%
   \label{mixing}
\end{figure}

\begin{figure}[h!]
     \begin{center}
            \includegraphics[width=0.48\textwidth]{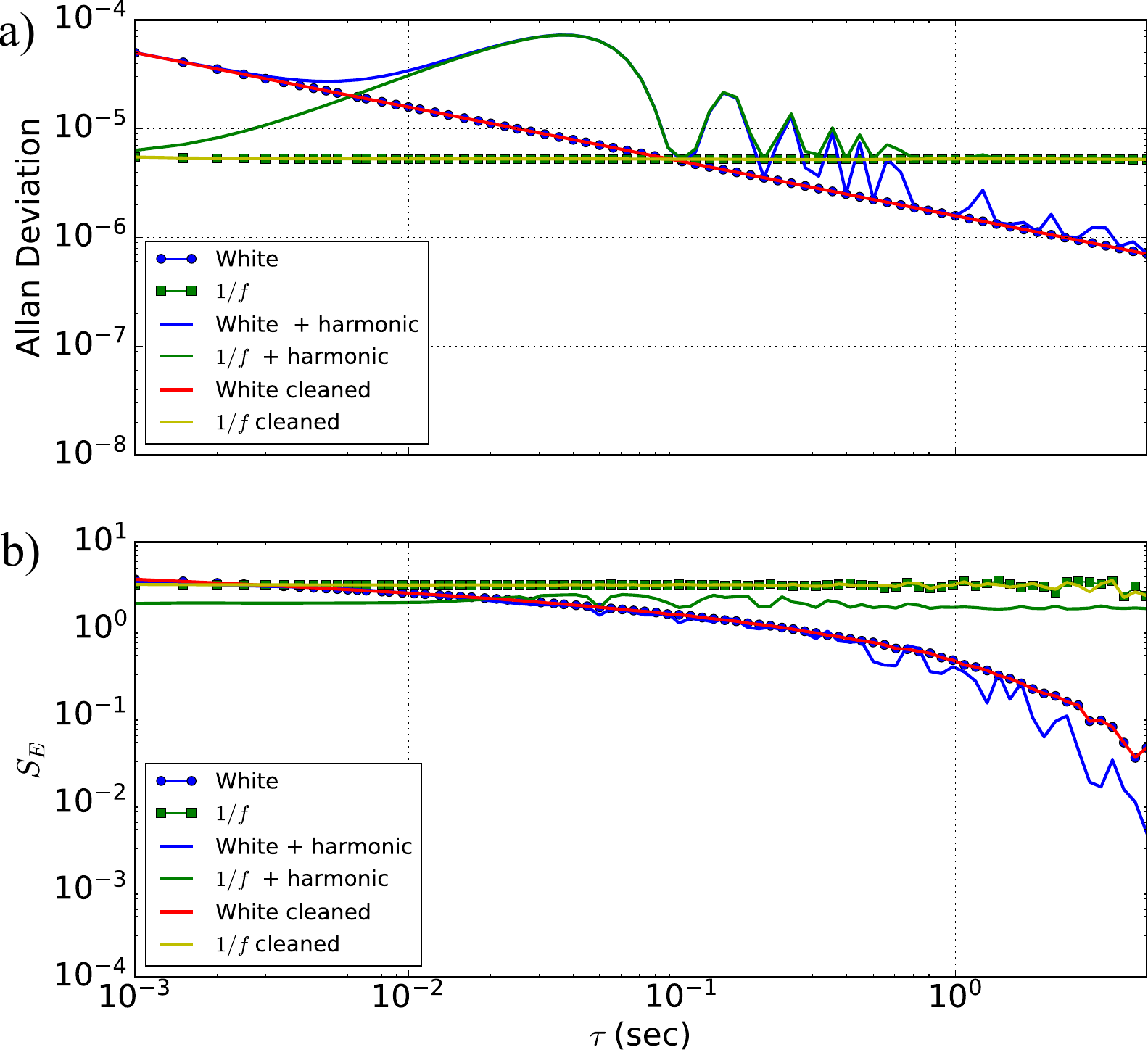}
            \end{center}
    \caption{Comparison of a) Allan Deviation and b) multi scale sample entropy with and without a strong spurious harmonic signal as well as after MAD filtering.}%
   \label{spurs}
\end{figure}

Replacing $v_\mathrm{white}(t)$ with a coherent harmonic tone and $S_E^{\mathrm{white}}$ with the corresponding multi scale sample entropy of that harmonic tone in Eq.~(\ref{eq:mixed}) and (\ref{FF0002G}), it can be seen that $S_E$ has a more difficult time differentiating a harmonic tone and a mixed event based signal when compared to the case of just white noise. The green trace in Fig.~\ref{mixing} highlights this difficulty, demonstrating the need for rigorous treatment of spurious harmonics in time series before the computation of multi scale entropy metrics.
Fig.~\ref{spurs} further demonstrates how a strong harmonic can distort Allan deviation as well as $S_E$ when added to white and flicker noise processes. In both cases, a periodic distortion pattern is observed across all time scales. To mitigate the effect of harmonics, a median absolute deviation filter (MAD) \cite{Dodge2010} can be employed. This filtering method is used to remove spurious harmonic components in the frequency domain by providing a robust threshold resistant to outliers. After computing the Fast Fourier Transform (FFT) of the signal, the absolute magnitudes of frequency components are analysed, and filtered as:
\begin{equation}
	\label{FF0003G}
\text{MAD} = \text{median}(|X_i - \text{median}(X)|)
\end{equation}
where $X_i$ represents the FFT magnitudes. Spurious frequencies are identified if their deviation from the median exceeds a scaled threshold:
\begin{equation}
	\label{FF0003G}
|X_i - \text{median}(X)| > \text{threshold} \times \text{MAD}
\end{equation}
Detected outliers are then set to zero, and the cleaned signal is reconstructed using the inverse FFT (IFFT). This method effectively suppresses isolated spectral artefacts caused by harmonics, digitisation errors, or electromagnetic interference while preserving the true signal components.
Comparing traces before and after filtering in Fig.~\ref{spurs}, it is clear that signals containing harmful high frequency harmonics can be effectively removed such that their ADEV and entropy measures are not corrupted. The result shows no traces of spurious signals in both the Allan deviation and Sample entropy plots. 

\section{Searching for Transient Effects from Cosmic Backgrounds}
 Recent investigations have reported rare high energy events in high frequency gravitational wave detectors made from similar devices to quartz OCXOs \cite{Goryachev:2021aa}, one such possible source for these events is thought to be high energy charged cosmic particles.  Additionally, previous theoretical calculations and experimental observations have suggested that cosmic rays could limit the performance of bar gravitational wave antennas, such as NAUTILUS \cite{Coccia:1995aa,Clay:1997aa,Astone:2002aa,Astone:2008aa}, by inducing excess noise. Furthermore, there remains an open question as to the origin of fundamental $1/f$ noise limits in quartz oscillators and resonators \cite{Gagnepain:1981aa,Moulton:1988aa}, which remains a critical factor in determining the ultimate stability and sensitivity of precision timekeeping and sensing applications.

These motivations provide the context for an experimental test using multi scale entropy metrics as well as ADEV on clock measurements in environments of differing cosmic background. Comparative measurements were thus conducted measuring the frequency performance of two Wenzel BTULN OCXOs \cite{WenzelBTULN}. Measurements were repeated both above ground, and 1 km deep underground in the low background environment provided by the cryogenic experimental laboratory for low-background Australian research (CELLAR), which is a deep underground facility hosted by the Stawell underground physics laboratory (SUPL)\cite{SUPL1}. SUPL is located in the Stawell gold mine 235 km from Melbourne, and provides the unique low background environment shielded from cosmic particles that is a requisite setting for various fundamental physics experiments\cite{ZUROWSKI2023,Fu:2026aa}.

\begin{figure}
     \begin{center}
            \includegraphics[width=0.39\textwidth]{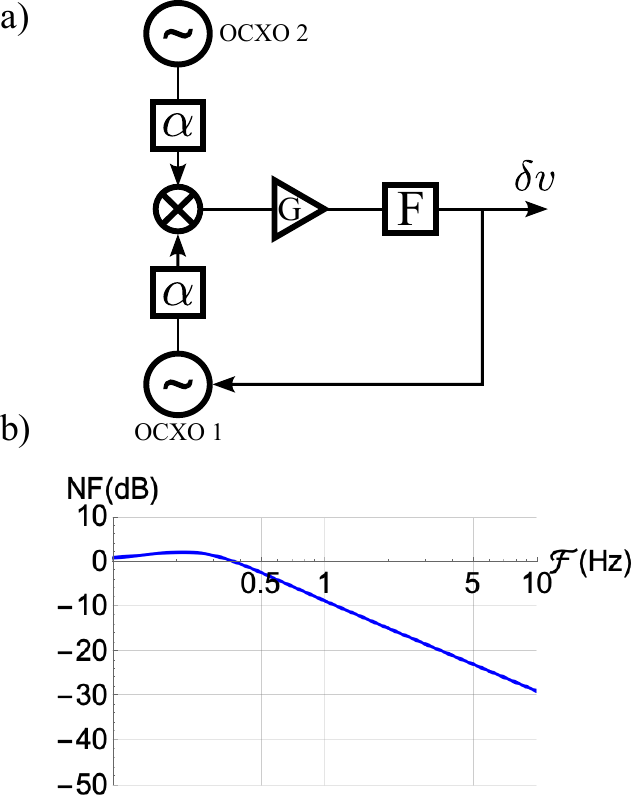}
            \end{center}
    \caption{\label{fig:setup} a) Measurement scheme of two OCXO frequency fluctuations. b) Noise factor NF transfer function that is applied by the closed-loop filter $\mathrm{F}$.}%
\end{figure}

In order to measure the performance of two OCXOs, their 10 MHz carrier signals were combined in a phase-locked loop (PLL) such that the frequency fluctuations of the combined signal can be accurately measured. Observing the configuration presented in figure \ref{fig:setup} a), The signal from the master oscillator (OCXO 2) is mixed with that of the slave (OCXO 1) after appropriate attenuation, and the mixer output voltage is then amplified and passed through a loop filter with transfer function given in figure \ref{fig:setup} b). The filtered correction signal is then fed back to the control port of the slave, so that the two oscillators are locked within the bandwidth of the PLL, but are free running outside this bandwidth. The correction voltage at the output of the filter $\delta v$ is sampled in the time domain to give the fractional frequency time series of the combined signal. Measurements were made utilising the exact same set up both underground at CELLAR, and above ground in Melbourne, with a portable power supply provided such that electrical noise contributions be as similar as possible. Each experiment was sampled for a total time of two hours at a rate $f_s=10$ kHz with an additional waiting period to ensure the oscillator ovens reached a stable temperature. Although noise at frequencies greater than the PLL bandwidth of $\approx$ 0.5 Hz will be largely suppressed, the oscillators are in a free running regime and fractional frequency noise can still be measured.
\begin{figure}
     \begin{center}
            \includegraphics[width=0.45\textwidth]{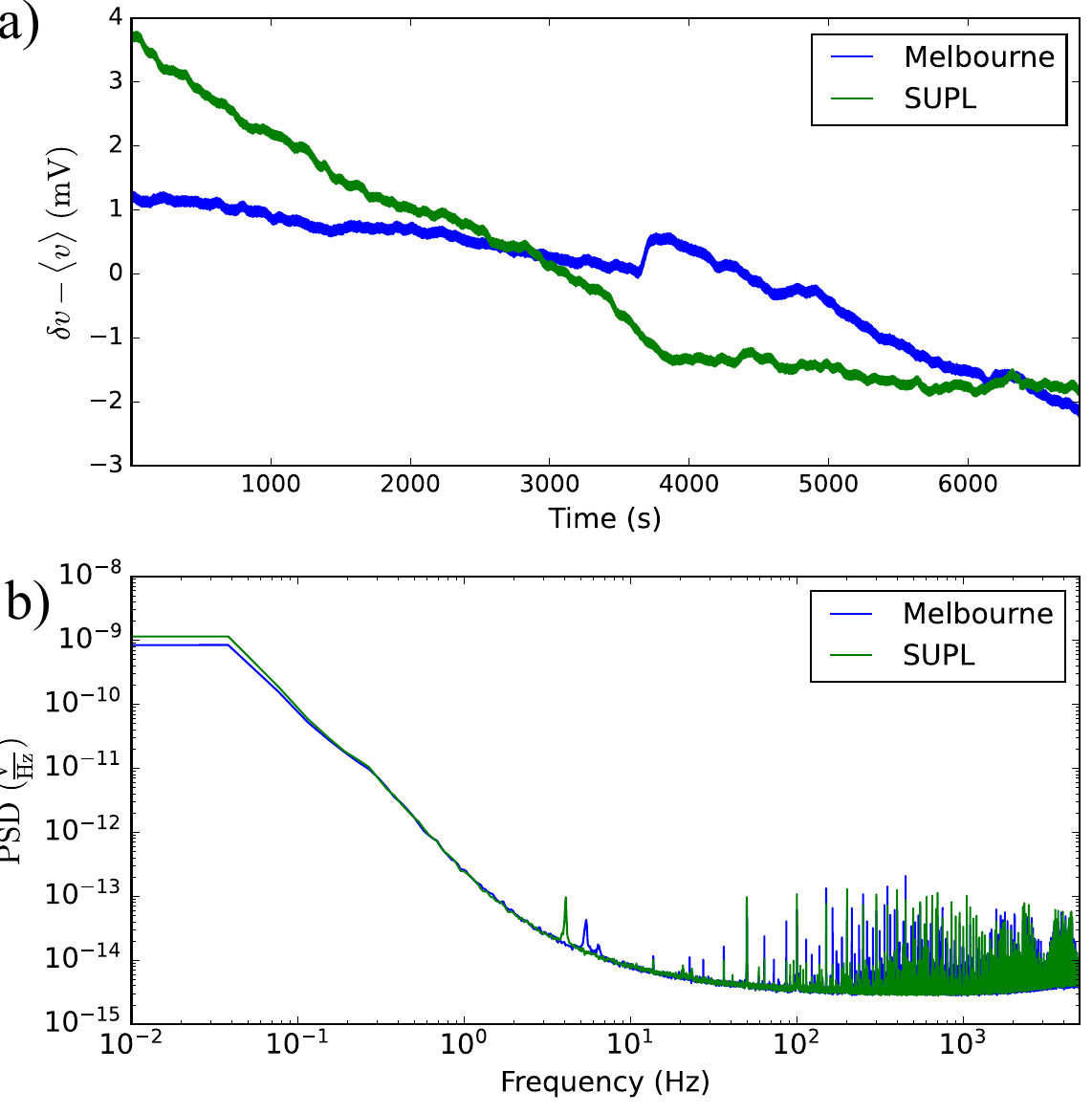}
            \end{center}
    \caption{\label{fig:time-series}a) Mixer correction voltage deviations from the mean level $\langle v \rangle$ for the oscillator signal both above (Melbourne) and below ground (SUPL). b) Power spectral density of above and below ground voltage signals.}%
\end{figure}

In figure \ref{fig:time-series} a), the PSD and mean subtracted time series of $\delta v$ is presented. Strong spurious harmonics at high frequencies can be observed in figure \ref{fig:time-series} a) that persist through the PLL filter transfer function. The fractional frequency variation is determined from $\delta v$ by $\delta f / f_c =\delta v / f_c \times df/dv $ where $df/dv$ is the tuning coefficient of the OCXO in units of HzV$^{-1}$ and $f_c$ is the carrier signal of 10 MHz. The inferred fractional frequency noise spectrum is then found by filtering away the PLL noise suppression factor $\mathrm{NF}$ of figure \ref{fig:setup} b).

\begin{figure}
	\begin{center}
		\includegraphics[width=0.45\textwidth]{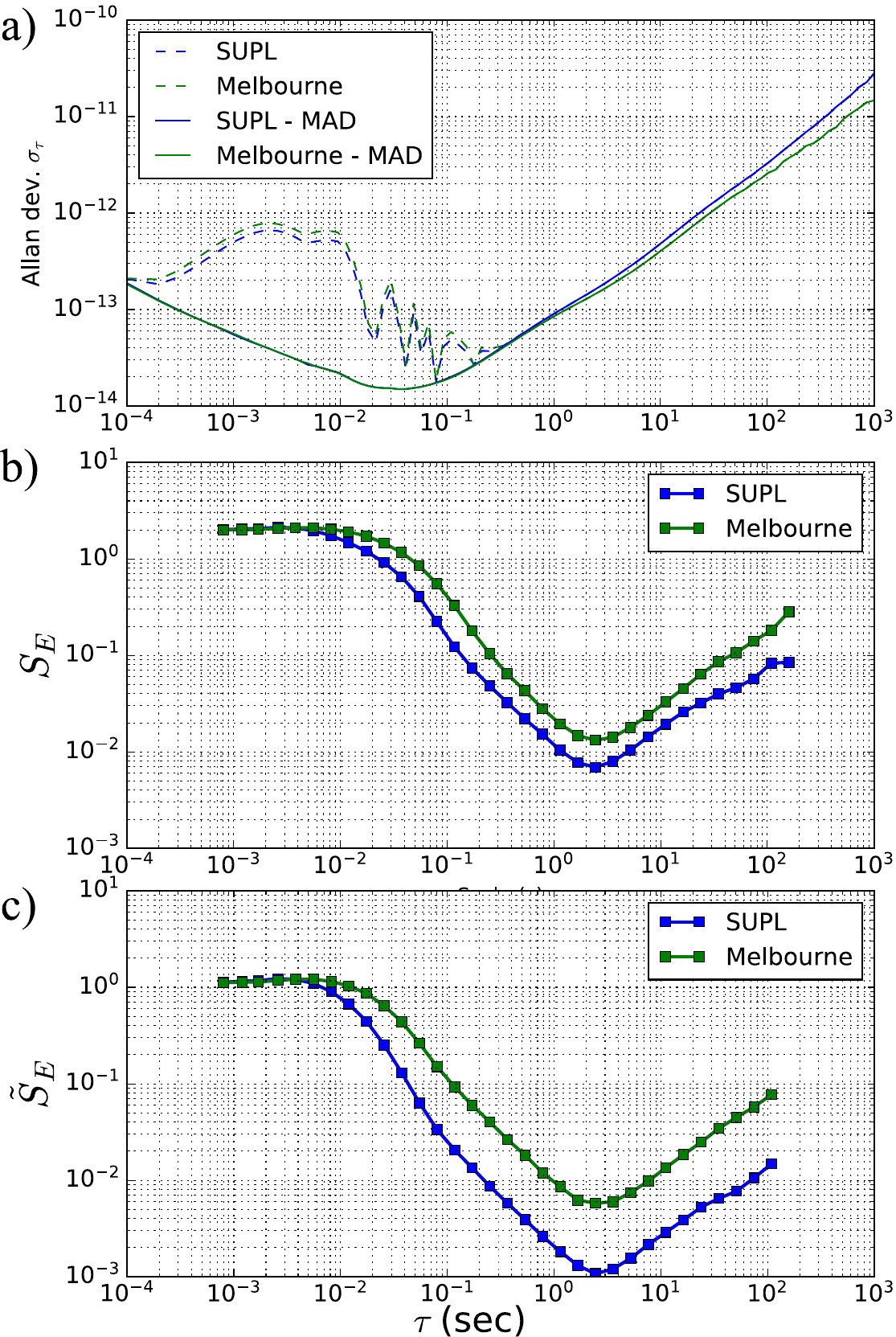}
	\end{center}
	\caption{\label{fig:entropy}Comparison of a) Allan deviation, b) sample entropy and c) modified sample entropy methods above (Melbourne) and below (SUPL) ground.}%
	
\end{figure}
The ADEV determined from $\delta f / f_c$ in both above and below ground experiments is presented as the dashed lines in figure \ref{fig:entropy} a). It is clearly observed that the high frequency harmonics distort the calculated variance at short time scales. MAD filtering was thus employed utilising the same threshold level on both data sets. As a result of the MAD procedure, large harmonics are successfully removed and the ADEV can be accurately recovered at short time scales, this can be seen as the solid lines in figure \ref{fig:entropy} a). No significant difference can be seen in the cleaned ADEV between both above and below ground measurements.

Multi-scale sample entropy $S_E$ as well as its modified counterpart $\tilde{S}_E$ were then calculated using the inferred fractional frequency noise as the input. The results as observed in figure \ref{fig:entropy} b) and c) demonstrate similar scale behaviour to the combined noise signal modelled in figure \ref{OCXO}, in that the entropy is dominated by $1/f$  and white noise uncertainties at short time scales, then dips in mid range integration times as white noise entropy contributions begin to decrease, until it ultimately rises again as coloured $1/f^n,~n>1$ noise contributions such as random walk noise lead to large unpredictability on longer scales.

A noticeable difference arises in both entropy metrics as the above-ground and underground measurements diverge with increasing time scale, with the underground data demonstrating a higher degree of predictability. This behaviour indicates a reduction in the entropy contribution associated with white-noise-like fluctuations as well as longer-correlated $1/f^n$ processes in the underground measurements. Based on the modelling presented above, changes associated purely with long-timescale thermal drift would be expected to appear primarily in the rising slope of $S_E$ and $\tilde{S}_E$, corresponding to slow noise components with exponent $n>2$. {The observed separation over the mid- to long-timescale range therefore indicates a broader change in the temporal predictability of the two frequency-fluctuation records, rather than only a simple instantaneous excitations.
 A muon crossing the oscillator package, resonator mount, electronics, or surrounding material can deposit energy impulsively and generate secondary excitations such as local heating, stress relaxation, acoustic ringing, charge redistribution, or perturbations of the sustaining and control electronics. In a high-$Q$ quartz oscillator, such disturbances need not appear only as isolated spikes: the resonator, oven, phase-locked loop, and surrounding thermal-mechanical structure can convert a short energy-deposition event into a slowly relaxing frequency perturbation or a change in the local correlation structure of the frequency time series. Reducing the muon flux underground would therefore be expected to reduce the occurrence of these intermittent relaxation events, making the oscillator record more repetitive and predictable at long coarse-graining times. This is precisely the type of change to which multi-scale sample entropy is sensitive, because it measures the recurrence of temporal patterns rather than only the variance of adjacent frequency averages. However, the entropy separation alone does not uniquely identify muons as the microscopic cause. It demonstrates an environment-dependent change in the temporal structure of OCXO frequency fluctuations. A definitive attribution to muons would require coincident particle detection, controlled shielding tests, or repeated measurements in environments where temperature, vibration, electromagnetic pickup, pressure, and other slow environmental variables are independently constrained. }

\section*{Conclusion}

Whilst the observed entropy difference between above-ground and underground datasets confirms an increase in the regularity of oscillator dominated frequency fluctuations in an underground environment, the physical interpretation of sample-entropy based observables is not yet uniquely determined. Sample entropy and its multi-scale generalizations quantify the statistics of pattern recurrence and predictability rather than a single low-order moment of the fluctuations. Consequently, multiple underlying mechanisms can in principle, produce similar shifts in $S_E$ and $\tilde{S}_E$. For example, a change in the rate or amplitude distribution of intermittent transients, a modification of correlation structure at long timescales, or alterations in the effective mixture of stationary and nonstationary components could all map onto an apparent change in multi-scale entropy. In this sense, the entropy difference should be viewed as a strong phenomenological discriminator between the two environments, but not as a definitive identification of a single microscopic noise source.

Irrespective of the ultimate mechanism, entropy metrics have been shown to provide complementary information to standard frequency-stability and spectral analyses, and can reveal differences that remain difficult to discern from Allan deviation and conventional power spectral density estimates. In our measurements, ADEV and spectral methods alone did not yield a statistically compelling separation between above-ground and underground operation over the explored range of averaging times, whereas $S_E$ and $\tilde{S}_E$ exhibited a pronounced divergence at longer effective integration scales. This demonstrates that the dominant distinction between the datasets is not primarily a change in the magnitude of second-order fluctuations, but rather a change in temporal structure. Moreover, the multi-scale framework enables identification of prevailing stochastic regime across time scales, providing a practical route to distinguishing flicker-like behavior from intermittent event-like processes.

These findings have direct implications for both fundamental-physics instrumentation and precision frequency metrology. Quartz BAW resonators are central components of proposed high-frequency gravitational-wave detectors and related resonant sensors, where the ability to distinguish weak, rare signals from non-Gaussian backgrounds is critical. In such experiments, environmental couplings that manifest as transients impulses, potentially sourced by cosmic radiation backgrounds, represent a limiting systematic and a potential source of false triggers. Our results indicate that multi-scale entropy can serve as a sensitive diagnostic for such nonstationary and rare-event contamination, offering a quantitative way to assess whether changes in environment or shielding alter the structure of noise rather than merely its variance. The same capability is valuable for frequency metrology more broadly, where long-term stability is frequently limited by poorly understood $1/f$-type processes and where traditional stability metrics may be insufficient in discriminating between heterogeneous or intermittent underlying processes.

This work was funded by the Australian Research Council Centre of Excellence for Engineered Quantum Systems, CE170100009 and Centre of Excellence for Dark Matter Particle Physics, CE200100008. Maxim Goryachev is supported by the Australian Research Council Future Fellowship. FT250100168. We would like to acknowledge Wenzel Associates, Inc for contributing the streamline OCXOs utilised in his work. Ben McAllister is supported by Australian Research Council grant DE250100933.

\hspace{10pt}

%\bibliography{biblioBAW}

\appendix

\section{Definition and Computation of Multi-Scale Sample Entropy}

\renewcommand{\theequation}{\thesection\arabic{equation}}

\setcounter{equation}{0}

\label{app:multiscale_sample_entropy}

This appendix gives the explicit definition of the sample-entropy metrics used in this work. Let
\begin{equation}
    x = \{x_i\}_{i=1}^{N}
\end{equation}
denote a uniformly sampled time series, taken here to be the fractional-frequency fluctuation record obtained from the oscillator readout. The sampling frequency is denoted by $f_s$, so that the sampling interval is $\Delta t = 1/f_s$.

\subsection{Coarse-grained time series}

To evaluate the regularity of the signal over different effective integration times, the original time series is first coarse grained. For an integer scale factor $M$, the coarse-grained series is defined as
\begin{equation}
    X^{(M)}_k =
    \frac{1}{M}
    \sum_{j=1}^{M}
    x_{(k-1)M+j},
    \qquad
    k = 1,\ldots,N_M,
    \label{eq:cg_series}
\end{equation}
where
\begin{equation}
    N_M = \left\lfloor \frac{N}{M} \right\rfloor .
    \label{eq:NM}
\end{equation}
The corresponding physical timescale is
\begin{equation}
    \tau = M \Delta t = \frac{M}{f_s}.
    \label{eq:tau_M}
\end{equation}
Thus, small $M$ probes short-timescale structure, while large $M$ probes the predictability of the frequency fluctuations after averaging over longer intervals.

\subsection{Sample entropy at a fixed scale}

For each coarse-grained sequence $X^{(M)} = \{X^{(M)}_k\}_{k=1}^{N_M}$, sample entropy is computed by comparing repeated patterns of length $m$ and $m+1$. The embedding dimension $m$ specifies the number of consecutive samples used to define a local temporal pattern. In this work we use $m=2$, so the algorithm tests whether pairs of neighbouring samples that are similar remain similar when extended to three samples.

For a given embedding dimension $m$, define the delay vectors
\begin{equation}
    \mathbf{X}^{(M,m)}_i =
    \left(
    X^{(M)}_i,
    X^{(M)}_{i+1},
    \ldots,
    X^{(M)}_{i+m-1}
    \right),
    \qquad
    i = 1,\ldots,N_M-m+1 .
    \label{eq:delay_vectors}
\end{equation}
The distance between two embedded vectors is taken to be the Chebyshev, or maximum-norm, distance
\begin{equation}
    d^{(m)}_{ij} =
    \max_{0\leq q < m}
    \left|
    X^{(M)}_{i+q} - X^{(M)}_{j+q}
    \right| .
    \label{eq:chebyshev_distance}
\end{equation}
Two patterns are considered matched if their distance is less than or equal to a tolerance $r$. Excluding self-matches, the number of matched pairs of length $m$ is
\begin{equation}
    B_m(M,r) =
    \sum_{i=1}^{N_M-m}
    \sum_{j=i+1}^{N_M-m+1}
    \Theta\!\left(r-d^{(m)}_{ij}\right),
    \label{eq:Bm}
\end{equation}
where $\Theta(y)$ is the Heaviside step function, equal to unity for $y\geq 0$ and zero otherwise. Similarly, the number of matched pairs of length $m+1$ is
\begin{equation}
    A_m(M,r) =
    \sum_{i=1}^{N_M-m-1}
    \sum_{j=i+1}^{N_M-m}
    \Theta\!\left(r-d^{(m+1)}_{ij}\right).
    \label{eq:Am}
\end{equation}
The sample entropy of the coarse-grained sequence is then
\begin{equation}
    \operatorname{SampEn}\!\left(m,r,X^{(M)}\right)
    =
    -\ln
    \left[
    \frac{A_m(M,r)}{B_m(M,r)}
    \right].
    \label{eq:sampen}
\end{equation}
Equivalently, sample entropy is the negative logarithm of the conditional probability that two sequences which match for $m$ consecutive samples also match when extended to $m+1$ consecutive samples. A smaller value therefore corresponds to a more predictable or more repetitive time series, while a larger value corresponds to reduced pattern recurrence and greater irregularity.

\subsection{Multi-scale sample entropy with fixed tolerance}

The first entropy metric used in this work, denoted $S_E$, is obtained by applying Eq.~\eqref{eq:sampen} at each coarse-graining scale $M$, while keeping the tolerance fixed relative to the standard deviation of the original time series. Specifically,
\begin{equation}
    S_E(\tau)
    =
    \operatorname{SampEn}\!\left(
    m,
    r_0,
    X^{(M)}
    \right),
    \label{eq:SE}
\end{equation}
where
\begin{equation}
    r_0 = \rho\,\sigma_x .
    \label{eq:r0}
\end{equation}
Here $\sigma_x$ is the standard deviation of the original uncoarse-grained record $x$, and $\rho$ is a dimensionless tolerance parameter. A common choice in sample-entropy analysis is $\rho=0.2$, with $m=2$. In this convention, the amplitude tolerance used to define pattern similarity is independent of coarse-graining scale. Therefore, changes in $S_E(\tau)$ reflect changes in the recurrence statistics of the coarse-grained signal under a fixed absolute similarity criterion.

\subsection{Modified multi-scale sample entropy with scale-dependent tolerance}

We also evaluate a modified multi-scale sample entropy, denoted $\widetilde{S}_E$, in which the tolerance is recomputed at each scale. This quantity is defined as
\begin{equation}
    \widetilde{S}_E(\tau)
    =
    \operatorname{SampEn}\!\left(
    m,
    r_M,
    X^{(M)}
    \right),
    \label{eq:SE_tilde}
\end{equation}
with
\begin{equation}
    r_M = \rho\,\sigma_{X^{(M)}} ,
    \label{eq:rM}
\end{equation}
where $\sigma_{X^{(M)}}$ is the standard deviation of the coarse-grained sequence at scale $M$. This normalization makes the matching threshold adaptive to the variance remaining after coarse graining. It can be useful for comparing pattern irregularity independently of the scale-dependent reduction or enhancement of signal amplitude. However, because the tolerance itself changes with $M$, the interpretation of $\widetilde{S}_E(\tau)$ differs from that of $S_E(\tau)$: the former measures recurrence statistics under a relative, scale-dependent similarity criterion, while the latter measures recurrence under a fixed tolerance inherited from the original record.

\subsection{Practical considerations}

For each value of $M$, Eqs.~\eqref{eq:Bm} and \eqref{eq:Am} require a sufficient number of embedded vectors to provide reliable counting statistics. Therefore, the largest usable scale is constrained by the record length through $N_M=\lfloor N/M\rfloor$. In practice, scales for which $A_m(M,r)=0$ or $B_m(M,r)=0$ are excluded or treated as undefined, since Eq.~\eqref{eq:sampen} would otherwise be singular.

The distinction between $S_E$ and $\widetilde{S}_E$ is important for interpreting oscillator data. The fixed-threshold entropy $S_E$ preserves information about both amplitude changes and temporal-pattern recurrence relative to the original fluctuation scale. The modified entropy $\widetilde{S}_E$, by contrast, partially removes the effect of scale-dependent variance and emphasizes changes in normalized temporal structure. Used together, these two quantities provide complementary diagnostics of oscillator fluctuations and can reveal intermittent or non-Gaussian structure that may not be apparent in Allan deviation or power spectral density estimates.

\end{document}